\documentclass[letter]{elsarticle}

\usepackage{amsmath}
\usepackage{amssymb}
\usepackage{anysize}
\usepackage{bold-extra}
\usepackage{color}
\usepackage{enumerate}
\usepackage{fancyhdr}
\usepackage{graphicx}
\usepackage[linktocpage,colorlinks]{hyperref}
\usepackage{lscape}
\usepackage{mathrsfs}
\usepackage{multirow}
\usepackage[final]{pdfpages}
\usepackage{rotating}
\usepackage{subfig}
\usepackage{textcomp}
\usepackage{url}
\usepackage{verbatim}
\usepackage{wrapfig}
\def\nn{\nonumber}

\begin{document}

\begin{frontmatter}

\title{Vector Leptoquarks and the 750 GeV Diphoton Resonance at the LHC}

\author{Christopher W. Murphy,}
\address{Scuola Normale Superiore, \\ Piazza dei Cavalieri 7, Pisa 56126, Italy}
\ead{christopher.murphy@sns.it}

\begin{abstract}
The ATLAS and CMS collaborations recently presented evidence of a resonance decaying to pairs of photons around 750 GeV. In addition, the BaBar, Belle, and LHCb collaborations have evidence of lepton non-universality in the semileptonic decays of $B$ mesons. In this work, we make a first step towards a unified explanation of these anomalies. Specifically, we extend the Standard Model by including vector leptoquarks and a scalar singlet that couples linearly to pairs of the leptoquarks. We find there is parameter space that gives the correct cross section for a putative 750 GeV resonance decaying to photons that is consistent with unitarity, measurements of the properties of the 125 GeV Higgs boson, and direct searches for resonances in other channels. In addition, we also show that constraints can be derived on any Beyond the Standard Model explanation of the 750 GeV resonance where the only new particles are scalars, which are strong enough to rule out certain types of models entirely. 
\end{abstract}

\end{frontmatter}

\section{Introduction}
The ATLAS and CMS collaborations recently presented evidence using data taken at $\sqrt{s} = 13$ TeV of a resonance, denoted $X$, decaying to a pair of photons~\cite{atlas:14110174, CMS:2015dxe}. ATLAS reported a local (global) significance of $3.6~\sigma \div 3.9~\sigma$ ($2.0~\sigma \div 2.3~\sigma$) for a mass $m_X \approx 750$ GeV. The range of significance depends on the assumptions made about the width of the resonance, with $3.6~\sigma$ corresponding to the narrow width approximation, and $3.9~\sigma$ when the width is taken to be about $6\%$ of the mass of the resonance. ATLAS recorded $8 \pm 4$ events above the expected background in the 750 GeV bin with $3.2~\text{fb}^{-1}$ of data. This translates to a cross section times branching ratio for the resonance of approximately $\sigma(p p \to X)~\text{Br}( X \to \gamma \gamma) \approx 1 \div 4$~\text{fb}. Meanwhile, CMS reported a local (global) significance of $2.6~\sigma$ ($1.2~\sigma$) for $m_X \approx 760$ GeV, consistent with ATLAS' findings. CMS assumed a narrow width for the resonance. While anomalies with local significances of $2~\sigma \div 3~\sigma$ are known to come and go,\footnote{For example, a preliminary CMS analysis of 7 \& 8 TeV data that searched for a resonance decaying to a pair of photons showed a $2.93~\sigma$ excess around $m_{\gamma\gamma} \approx 136.5$ GeV~\cite{CMS:2013wda}. However, this excess disappeared in the final version of the analysis~\cite{Khachatryan:2014ira}.} the fact that both CMS and ATLAS see an excess around the same mass has already led to considerable theoretical attention~\cite{Harigaya:2015ezk, Mambrini:2015wyu, Backovic:2015fnp, Angelescu:2015uiz, Nakai:2015ptz, Knapen:2015dap, Buttazzo:2015txu, Pilaftsis:2015ycr, Franceschini:2015kwy, DiChiara:2015vdm, Higaki:2015jag, McDermott:2015sck, Ellis:2015oso, Low:2015qep, Bellazzini:2015nxw, Gupta:2015zzs, Petersson:2015mkr, Molinaro:2015cwg, Dutta:2015wqh, Cao:2015pto, Matsuzaki:2015che, Kobakhidze:2015ldh, Martinez:2015kmn, Cox:2015ckc, Becirevic:2015fmu, No:2015bsn, Demidov:2015zqn, Chao:2015ttq, Fichet:2015vvy, Curtin:2015jcv, Bian:2015kjt, Chakrabortty:2015hff, Ahmed:2015uqt, Agrawal:2015dbf, Csaki:2015vek, Falkowski:2015swt, Aloni:2015mxa, Bai:2015nbs, Gabrielli:2015dhk, Benbrik:2015fyz, Kim:2015ron, Alves:2015jgx, Megias:2015ory, Carpenter:2015ucu, Bernon:2015abk, Chao:2015nsm, Arun:2015ubr, Ding:2015rxx, Han:2015qqj, Cao:2015twy, Heckman:2015kqk, Bauer:2015boy, Barducci:2015gtd, Huang:2015rkj, Gu:2015lxj, Davoudiasl:2015cuo, Son:2015vfl, Goertz:2015nkp, Kanemura:2015vcb, Hernandez:2015hrt, Danielsson:2016nyy, Modak:2016ung, Fabbrichesi:2016alj, Hati:2016thk, Dorsner:2016ypw}. See~\cite{Jaeckel:2012yz, Grinstein:2013fia} for some studies of the prospect of resonances in the diphoton channel using Run-1 LHC data.

To make things even more exciting, for the past couple of years there have been several anomalies in the semileptonic decays of $B$ mesons. In particular, the BaBar, Belle, and LHCb collaborations have seen evidence for a deviation from $\tau / \ell$ universality (where $\ell = e, \mu$) in the semileptonic, charged current decays $B \to D^{(*)}$,
\begin{equation}
R^{\tau/\ell}_{D^{(*)}} = \frac{\text{Br}(\bar{B} \to D^{(*)} \tau \bar{\nu}) / \text{Br}(\bar{B} \to D^{(*)} \tau \bar{\nu})_{SM}}{\text{Br}(\bar{B} \to D^{(*)} \ell \bar{\nu}) / \text{Br}(\bar{B} \to D^{(*)} \ell \bar{\nu})_{SM}}.
\end{equation}
Combining the three experiments' measurements, $R^{\tau/\ell}_{D} = 1.37 \pm 0.17$, $R^{\tau/\ell}_{D^*} = 1.28 \pm 0.08$, leads to a departure from the Standard Model (SM) prediction at the $3.9~\sigma$ level~\cite{Lees:2013uzd, Huschle:2015rga, Aaij:2015yra}. In addition, LHCb has $2.6~\sigma$ evidence of a departure from $\mu / e$ universality in the neutral current $b \to s$ decay,
\begin{equation}
R^{\mu / e}_K = \frac{\text{Br}(B \to K \mu^+ \mu^-)}{\text{Br}(B \to K e^+ e^-)} = 0.745^{+0.090}_{-0.074} \pm 0.036,
\end{equation}
which is predicted to be very close to one in the SM~\cite{Aaij:2014ora}. These anomalies have also attracted significant theoretical attention, see e.g.~\cite{Datta:2013kja, Alonso:2014csa, Hiller:2014yaa, Altmannshofer:2014rta, Gripaios:2014tna, Bhattacharya:2014wla, Sahoo:2015wya, Sierra:2015fma, Becirevic:2015asa, Alonso:2015sja, Greljo:2015mma, Calibbi:2015kma, Freytsis:2015qca, Descotes-Genon:2015uva, Bauer:2015knc, Fajfer:2015ycq, Barbieri:2015yvd}.

The primary model building challenge associated with the 750 GeV resonance is to explain why the decay of $X$ to a pair of photons, which is a loop level process for the Higgs boson of the SM, was the first channel in which evidence of a resonance showed up. In particular, the cross section times branching ratio for $p p \to X \to \gamma\gamma$ is $O(10^4)$ times larger than that of a 750 GeV Higgs boson in the SM~\cite{Heinemeyer:2013tqa}. Since $X$ is neutral, it must couple to pairs of charged particles in order to decay to a pair of photons. Furthermore, the most straightforward to produce a neutral scalar at a hadron collider is through gluon fusion, which will occur if the scalar is coupled to pairs of colored particles . Many of the proposed explanations for the $B$ decay anomalies involve leptoquarks, which are charged and colored bosons. The preferred mass of the leptoquarks is $O(1~\text{TeV})$ with couplings of $O(1)$ to quarks and leptons, see e.g.~\cite{Alonso:2015sja, Greljo:2015mma, Barbieri:2015yvd}. 

With this background, it would seem that an electroweak (EW) scalar singlet coupled linearly to a pair of leptoquarks should be a good candidate to explain the 750 GeV resonance, while simultaneously explaining the $B$ decay anomalies. Several authors have explored the non-minimal scenario where there are scalar leptoquarks, and additional particles in the loops, such as vector-like fermions~\cite{Bauer:2015boy, Hati:2016thk, Dorsner:2016ypw}. Refs.~\cite{Bauer:2015boy, Hati:2016thk} also noted the potential for their models to explain the $B$ decay anomalies.\footnote{Leptoquarks are also viable explanations of the CMS $\ell \ell j j$ anomaly~\cite{Khachatryan:2015vaa}, see e.g.~\cite{Hati:2016thk, Queiroz:2014pra, Allanach:2015ria}.} In contrast with those works, we study a model involving a scalar singlet and vector leptoquarks.

The rest of the paper is organized as follows. We argue in Sec.~\ref{sec:arg} that a Beyond the Standard Model (BSM) theory where the only new particles are scalars, and whose interactions arise purely from dimension-4 operators, cannot be an explanation of the resonance $X$. We also show that similar logic can be used to constrain, but not in general completely rule out both extended scalar sectors with interactions arising from both dimension-3 and dimension-4 operators, and theories with new spin-1 particles. In Sec.~\ref{sec:vec}, we study a model involving a scalar singlet and vector leptoquarks. Since there are already multiple works which show the viability of vector leptoquarks as explanations of $R^{\tau/\ell}_{D^{(*)}}$ and $R^{\mu / e}_K$, we simply assume the leptoquarks are viable with $M_V \sim 1$ TeV, and focus solely on explaining the diphoton excess and the constraints directly associated with it. In this initial work, we ignore the novel loop level contributions in this model to $X \to f \bar{f}$ coming from a loop containing two leptoquarks and one fermion. We find there is parameter space that gives the correct cross section for a putative 750 GeV resonance decaying to photons that is consistent with unitarity, measurements of the properties of the 125 GeV Higgs boson, and direct searches for resonances in other channels.

\section{Theoretical Constraints on Explanations of the 750 GeV Resonance}
\label{sec:arg}

Ref.~\cite{Angelescu:2015uiz} found that the two-Higgs doublet model (2HDM) was not a viable explanation of the 750 GeV resonance. We find that the doublet-septet model of Ref.~\cite{Hisano:2013sn} cannot explain the excess. \textit{A priori}, the doublet-septet model was an attractive candidate as it does not contribute to the $\rho$ parameter at tree level~\cite{Veltman:1977kh, Agashe:2014kda}, and has six charged scalars with $Q = -1, 1, 2, 3, 4, 5$ to increase the diphoton rate. Since this failure seems to be a generic feature of models where the only BSM particles are scalars whose interactions arise purely from dimension-4 operators, it is worthwhile to have a more model-independent understanding of why this class of theories does not work. We give such an argument in what follows. Similar logic can be applied to the case where there are interactions mediated by both dimension-3 and dimension-4 operators, as well as to the case where there are spin-1 particles. We show that while models are constrained, there is in general parameter space that allows for an explanation of the diphoton excess. 

First, consider the scenario where all of the BSM particles are scalars, and all of the interactions in the scalar potential are due to dimension-4 operators. In this case, all of the neutral, $CP$-even scalars (and crucially in some theories, some of the charged scalars) are Higgs boson-like in the sense that they all have a cubic interaction involving one Higgs boson and a pair of EW vector bosons. The couplings of Higgs bosons to fermions and vector bosons are related to each other by sum rules~\cite{Grinstein:2013fia, Lee:1977yc, Lee:1977eg, Gunion:1990kf}. The sum rules arise by requiring the scattering of longitudinally polarized vector bosons to be unitary (at tree level) to arbitrarily high energies. The mixing angles that enter into the sum rules for Higgs bosons also appear in every cubic scalar coupling that arises purely from fields related to electroweak (EW) symmetry breaking. The 2HDM is known to have this property. Refs.~\cite{Bernon:2015qea, Bernon:2015wef} showed that the cubic coupling of $h$ to the charged Higgs boson in the 2HDM approach a finite, non-zero value in the alignment limit, which is the direction measurements from the LHC are pushing the parameter space of the 2HDM. We have verified that the cubic coupling of the charged Higgs boson to the would-be $X$ in the 2HDM (the heavier, $CP$-even Higgs boson) approaches zero. Furthermore, the couplings of the 125 GeV Higgs boson to fermions and vector bosons are Standard Model-like (SM-like) to a good extent~\cite{Aad:2012tfa, Chatrchyan:2012xdj, Aad:2015zhl, Khachatryan:2014jba, Aad:2015gba}.  Thus, any couplings related to those of the 125 GeV Higgs, $h$, must correspondingly be small, including the aforementioned cubic scalar couplings. 

Now consider the case where all the BSM particles are still scalars, but the scalar potential has both dimension-3 and dimension-4 operators. A simple way to generate a dimension-3 operator is to include a scalar singlet in the theory. There are theories without scalar singlets that have dimension-3 operators in their potential as well, such as the Georgi-Machacek model~\cite{Georgi:1985nv, Logan:2015xpa}. Cubic scalar interactions can evade the argument of the previous paragraph as they do not necessarily involve the mixing angles associated with EW symmetry breaking (e.g. $\beta$ in the 2HDM). Another advantage of dimension-3 operators, being relevant operators, is that they do not run to Landau poles. Indeed for the fairly large values of couplings needed in theories where dimension-4 operators are responsible for the diphoton excess, the quartic couplings run to Landau poles, sometimes at unacceptably low scales~\cite{Franceschini:2015kwy, Gu:2015lxj, Son:2015vfl, Goertz:2015nkp}. In models where dimension-3 operators are primarily responsible for the diphoton excess, the coefficents of the dimension-4 operators can simply be taken to be small enough such that the Landau pole scale is well above the other scales of interest in the problem.

There are two challenges in the case of dimension-3 operators. First, there is still a relation in these theories between the couplings $h$ and $X$ to pairs of SM particles. Both the relevant mixing angle(s) between $h$ and $X$ (e.g. $\alpha$ in the 2HDM) and the magnitude of the cubic coupling must be small enough such that the predictions for the signal strengths of the 125 GeV Higgs boson are consistent with experiment. 

The second consideration is that the magnitude of the cubic coupling must be below the maximum value that is theoretically allowed by perturbative unitarity~\cite{Lee:1977yc, Lee:1977eg, Dawson:1988va, Grinstein:2015rtl}. The perturbative unitarity bounds are most restrictive for scalars in the diphoton loop, as the scalar-loop form factor is the smallest of the three (scalar, fermion, vector), $A_0 \sim 1 / 3$ versus $A_{1/2} \sim 4 / 3$ and $A_1 \sim 7$~\cite{Gunion:1989we, Djouadi:2005gj}. These form factors are discussed further in Sec.~\ref{sec:vec}. Interactions of the form $X \phi^{Q+} \phi^{Q-}$ are bounded by $\phi^{Q+} \phi^{Q+} \to \phi^{Q+} \phi^{Q+}$ scattering.\footnote{We consider $\phi^{Q+} \phi^{Q+}$ rather than $\phi^{Q+} \phi^{Q-}$ scattering to avoid $s$-channel resonances (assuming $\phi^{Q+}$ is the scalar with the largest electric charge that couples to $X$), which complicates the process of extracting unitarity bounds~\cite{Dawson:1988va}.} A generic scattering amplitude of this type can be written as,
\begin{equation}
\label{eq:amp}
\mathcal{A} = - \lambda - m^2 \left(\frac{1}{t - M_X^2} + \frac{1}{u - M_X^2}\right),
\end{equation}
where $\lambda$ is a linear combination of quartic couplings, $m$ is a linear combination of cubic couplings, and we have assumed that the mixing between $X$ and $h$ is small enough to neglect the contribution of $h$. $|m| \gg v \approx 246$~GeV is needed to get the correct cross section to explain the diphoton excess, so $m$ should be mostly composed of couplings that arise from dimension-3 operators, otherwise the 2HDM would be viable. The strongest unitarity bounds on $\lambda$ and $m$ come from the $\ell = 0$ partial wave amplitude for the scattering amplitude, Eq.~\eqref{eq:amp}, evaluated at large $s$ and the threshold for on-shell $\phi^{Q+}$ scattering respectively,
\begin{equation}
16 \pi a_0 = 
\begin{cases}
- \lambda & \quad s \gg M_{X,Q} \\
\frac{2 m^2}{M_X^2} \frac{M_X^2 + 2 M_Q^2}{M_X^2 + 4 M_Q^2} - \lambda & \quad s \to 4 M_Q^2
\end{cases},
\end{equation}
with $M_Q$ being the mass of $\phi^{Q+}$. The constraint $|a_0| < 1\, (|\text{Re}(a_0)| < 1 / 2)$ at large $s$ leads to the bound $|\lambda| < 16 \pi\, (8 \pi)$. Plugging in the positive upper limit on $\lambda$ leads to the minimal, but model-independent bounds, $|m| < 6.2 \div 7.3$~TeV ($4.4 \div 5.1$~TeV) for $M_Q = 400 \div 1400$~GeV, again using the constraint $|a_0| < 1\, (|\text{Re}(a_0)| < 1 / 2)$. 

We now apply these bounds to some of the models that have recently been considered in the literature. However, due to large amount of literature on this subject, it is likely that we are not considering every relevant model. Ref.~\cite{Cao:2015twy} considers the singlet extension of the Manohar-Wise model~\cite{Manohar:2006ga}. Its benchmark point with $m_{S_i} = 1$~TeV is in trouble with unitarity as $m = k_2 f = 6.36$~TeV.  This is ruled out by $|\text{Re}(a_0)| < 1 / 2$, but allowed by $|a_0| < 1$. On the other hand, its benchmark point with $m_{S_i} = 600$~GeV is consistent with this unitarity analysis. Ref.~\cite{Kanemura:2015vcb} also looked at extended scalar sectors. They showed that with enough charged scalars the correct production rate for $X$ can be obtained. However, they do not indicate if this can be done without spoiling the agreement between theory and experiment for the coupling of the 125~GeV Higgs boson to photons. Ref.~\cite{Fabbrichesi:2016alj} does a parameter scan to find viable parameter space in the Georgi-Machacek model. In this case, the parameter of interest is $m = g_{H H_5^0 H_5^0} \sim 2 \sqrt{3} M_2$ for $M_2 \gg v$, $|\alpha| \approx 0$. The unitarity bound roughly leads to $|M_2| < 1.9\, (1.3)$~TeV for $m_{H_5} = 400 \div 600$~GeV using $|a_0| < 1\, (|\text{Re}(a_0)| < 1 / 2)$. $M_2$ is scanned over the range $1 \div 10^4$ GeV in Ref.~\cite{Fabbrichesi:2016alj}. 

Finally, consider the case where the scalar resonance, $X$, couples linearly to pairs of charged and/or colored, spin-1 particles. At high-energies, when these vector particles are longitudinally polarized, $|\epsilon_{\mu}| \sim \sqrt{s} / M_V$, the amplitude for elastic vector-vector scattering mediated by the exchange of $X$ grows with energy. At sufficiently high energies, unitarity will break down unless there is a cancellation amongst the various parameters that enter into the scattering amplitude, which we will assume does not happen. For a cubic coupling $2\kappa M_V^2 / v$, the amplitude schematically has the form,
\begin{equation}
\label{eq:vecU}
\mathcal{A} \sim \left(\frac{\sqrt{s}}{M_V}\right)^4 \left(\frac{\kappa M_V^2}{v}\right)^2 \frac{1}{s}, \quad a_0 \sim \frac{\kappa^2}{16 \pi} \frac{s}{v^2}, \quad \text{for}\, s \gg M_{X,V}^2,
\end{equation}
plus terms with less powers of energy in the numerator. Provided there is no cancellation among parameters, the theory break down around energies $\sqrt{s} \sim 4 \sqrt{\pi} v / | \kappa |$.

\section{Model of a Scalar Singlet and Vector Leptoquarks}
\label{sec:vec}
Consider supplementing the SM with a scalar singlet, $\phi$, and a vector leptoquark, $V_{\mu}$. All of the SM charge assignments for the various vector leptoquark bosons, and the current(s) of SM fermions that they couple to, are given in Tab.~\ref{tab:vec}.
\begin{table}
\centering
 \begin{tabular}{c c c | c}
 $SU(3)_c$ & $SU(2)_L$ & $U(1)_Y$ & $J_{\mu}$ \\ \hline 
 $\bar{3}$ & $1$ & $-2/3$ & $\bar{\ell}_L \gamma_{\mu} q_L$, $\bar{e}_R \gamma_{\mu} d_R$ \\
 $\bar{3}$ & $3$ & $-2/3$ & $\bar{\ell}_L \gamma_{\mu} q_L$ \\
 $\bar{3}$ & $2$ & $5/6$ & $\bar{\ell}_L^c \gamma_{\mu} d_R$, $\bar{e}_R^c \gamma_{\mu} q_L$ \\
 $\bar{3}$ & $1$ & $-5/3$ & $\bar{e}_R \gamma_{\mu} u_R$ \\
 $\bar{3}$ & $2$ & $-1/6$ & $\bar{\ell}_L^c \gamma_{\mu} u_R$
  \end{tabular}
  \caption{SM charge assignment for the various types of vector leptoquarks, $V^{\mu}$, and the SM currents of fermions they couple to. Adapted from Ref.~\cite{Alonso:2015sja}.}
  \label{tab:vec}
\end{table}
The interactions of interest are,
\begin{equation}
\label{eq:int}
\mathcal{L}_{int} = \kappa \frac{2 M_V^2}{v} \phi V_{\mu}^{\dagger} V^{\mu} + g \frac{2 M_V^2}{v^2}\left(H^{\dagger} H - \frac{v^2}{2}\right) V_{\mu}^{\dagger} V^{\mu},
\end{equation}
where $H$ is the Higgs doublet ($H^T = (0 , (v + h_2)/\sqrt{2})$ in unitary gauge), $M_V$ is the mass of the leptoquark, and $\kappa$, and $g$ are free parameters. We chose the parameterization of Eq.~\eqref{eq:int} for later convenience. Since there are already multiple works which show the viability of vector leptoquarks as explanations of $R^{\tau/\ell}_{D^{(*)}}$ and $R^{\mu / e}_K$, see for example Refs.~\cite{Alonso:2015sja, Greljo:2015mma, Barbieri:2015yvd}, we simply assume the leptoquarks are viable with $M_V \sim 1$ TeV, and focus solely on explaining the diphoton excess and the constraints directly associated with it.

The scalar singlet mixes with the neutral, $CP$-even component of the Higgs doublet, \begin{equation}
\label{eq:mat}
\begin{pmatrix}
\phi \\
h_2
\end{pmatrix} =
\begin{pmatrix}
\cos \alpha & - \sin \alpha \\
\sin \alpha & \cos \alpha
\end{pmatrix} 
\begin{pmatrix}
X \\
h
\end{pmatrix}.
\end{equation}
This mixing leads to a universal form for the decay rates of $X$ into SM particles that couple of $X$ at tree level,
\begin{equation}
\frac{\Gamma(X \to f \bar{f})}{\Gamma(X \to f \bar{f})_{SM}} = \frac{\Gamma(X \to V V)}{\Gamma(X \to V V)_{SM}} = \sin^2 \alpha,
\end{equation}
where $\Gamma(X \to Y)_{SM}$ is the branching ratio for a SM Higgs boson with $M_X = 750$ GeV, $VV = W^+W^-,\, ZZ$, and $f = t,\, b,\, \tau$. The mixing in Eq.~\eqref{eq:mat} cannot be too large, as it is constrained by measurements of $h$~\cite{Khachatryan:2014jba, Aad:2015gba}. Requiring all of the universal-type signal strengths for $h$ to be within 15\% of their SM values yields the bound $\alpha < 0.4$. We neglect the possibility of the tree level decay $X \to h h$, as the parameter that controls the rate of this decay can be made sufficiently small without affecting the interesting features of the model. The decay rates of $X$ into SM particles that do not directly couple to $X$ are more complicated,
\begin{align}
\label{eq:loop}
\frac{\Gamma(X \to g g)}{\Gamma(X \to g g)_{SM}} &= \left|\sin\alpha + (\kappa \cos\alpha + g \sin\alpha) \frac{A_1(\tau_{V})}{A_{1/2}(\tau_{t}) + A_{1/2}(\tau_{b})}\right|^2, \\
\frac{\Gamma(X \to \gamma \gamma)}{\Gamma(X \to \gamma \gamma)_{SM}} &= \left|\sin\alpha + (\kappa \cos\alpha + g \sin\alpha) \frac{N_{CV} Q_V^2 A_1(\tau_{V})}{\sum_f N_{Cf} Q_f^2 A_{1/2}(\tau_{f}) + A_1(\tau_{W})}\right|^2, \nn
\end{align}
with $\tau_a = M_X^2 / 4 M_a^2$. There is a similar formula for $X \to Z \gamma$ as well. The formulas for $h \to \{g g, \gamma \gamma, Z \gamma, Z Z, W W\}$ have a similar structure to Eqs.~\eqref{eq:loop} except the angular dependences of the SM and leptoquark contributions are $\cos\alpha$ and $-\kappa \sin\alpha + g \cos\alpha$ respectively. In addition, $M_h$ enters the form factors instead of $M_X$. From this, we see that measurements of the Higgs' couplings constrain both $\alpha$ and $g$ to be small. This illustrated explicitly in Figure~\ref{fig:vec} below. In addition, small $g$ is favored by constraints from running the quartic couplings of the theory.

In this initial work, we ignore the novel loop level contributions in this model to $X \to f \bar{f}$ coming from a loop containing two leptoquarks and one fermion, e.g. $X \to b \bar{b}$ receives a contribution at the one-loop level from a loop consisting of two leptoquarks and a tau. The form factors entering the one-loop amplitudes are~\cite{Gunion:1989we, Djouadi:2005gj},
\begin{align}
A_0(\tau) &= - [\tau - f(\tau)] \tau^{-2}, \\
A_{1/2}(\tau) &= 2 [\tau + (\tau - 1) f(\tau)] \tau^{-2}, \nn \\
A_1(\tau) &= - [2\tau^2 + 3 \tau + 3 (2 \tau - 1) f(\tau)] \tau^{-2}. \nn
\end{align}
$A_0$ is listed even though it does not appear in Eq.~\eqref{eq:loop} to help facilitate the discussion of Sec.~\ref{sec:arg}. The function $f(\tau)$ is,
\begin{equation}
f(\tau) = 
\begin{cases}
\arcsin^2 \sqrt{\tau} & \quad \text{if } \tau \leq 1 \\
- \frac{1}{4}\left(\ln \frac{1 + \sqrt{1 - 1 / \tau}}{1 - \sqrt{1 - 1 / \tau}} - i \pi\right)^2 & \quad \text{if } \tau > 1
\end{cases}.
\end{equation}

The cross section times branching ratio for $p p \to X \to \gamma \gamma$ is,
\begin{align}
\sigma(p p \to X) &= \sigma(ggF \to X)_{SM} \frac{\Gamma(X \to g g)}{\Gamma(X \to g g)_{SM}} + \sigma(VBF \to X)_{SM} \frac{\Gamma(X \to V V)}{\Gamma(X \to V V)_{SM}} , \\
\text{Br}( X \to \gamma \gamma) &= \text{Br}(X \to \gamma \gamma)_{SM} \frac{\Gamma(X \to \gamma \gamma)}{\Gamma(X \to \gamma \gamma)_{SM}} \left(\sum_Y \text{Br}(X \to Y)_{SM} \frac{\Gamma(X \to Y)}{\Gamma(X \to Y)_{SM}}\right)^{-1}, \nn
\end{align}
where $ggF$ and $VBF$ are the gluon and vector boson fusion production processes respectively, and $Y = t \bar{t},\, b \bar{b},\, \tau \bar{\tau},\, W^+ W^-,\, Z Z,\, g g,\, \gamma \gamma,\, Z \gamma$. All SM Higgs boson cross sections and branching ratios are taken from~\cite{Heinemeyer:2013tqa}. In this initial work, we neglect the possibility of producing the resonance through photon fusion. Including photon fusion would decrease the value of $\kappa$ needed to explain the excess, allowing more parameter space to satisfy the bounds from unitarity. 

\begin{figure}
  \centering
  \subfloat[$(\bar{3}, 3)_{-2/3}$, $\kappa = 0.10$]{\includegraphics[width=0.48\textwidth]{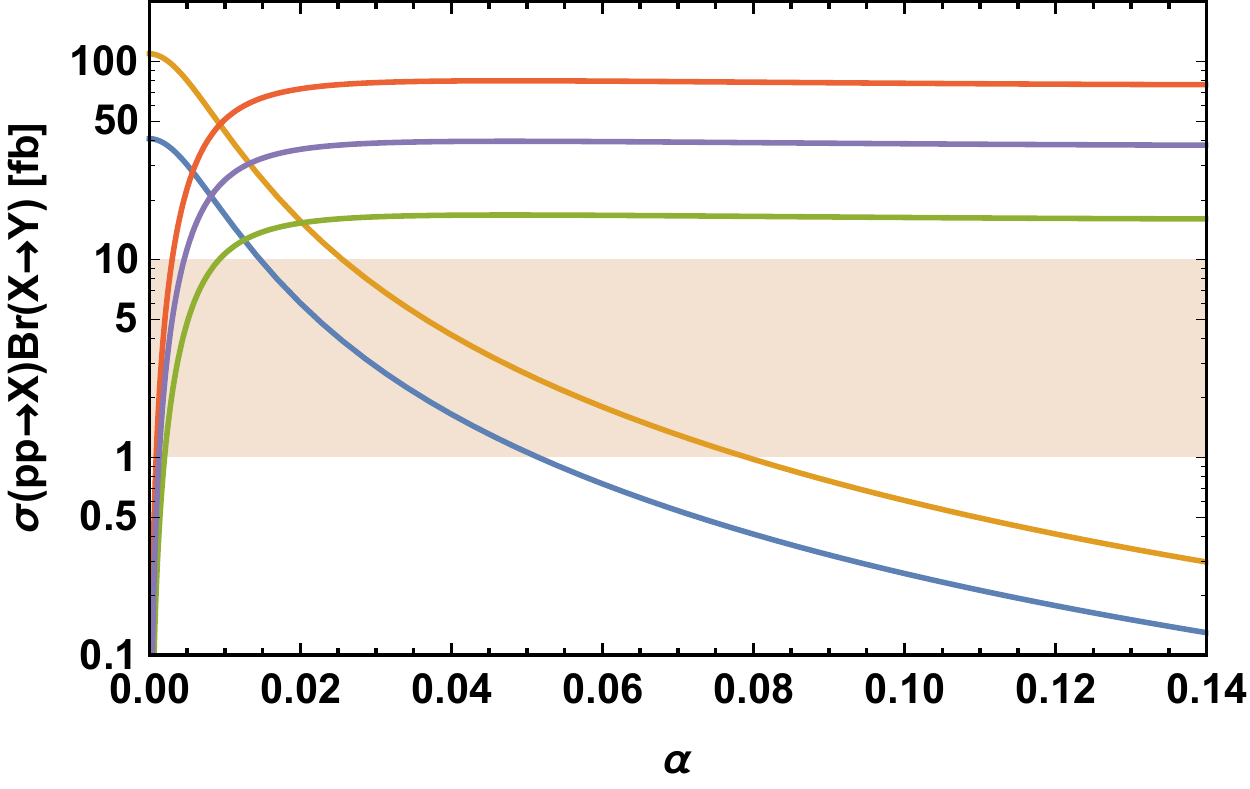}}\,
\subfloat[$(\bar{3}, 1)_{-2/3}$, $\kappa = 0.27$]{\includegraphics[width=0.48\textwidth]{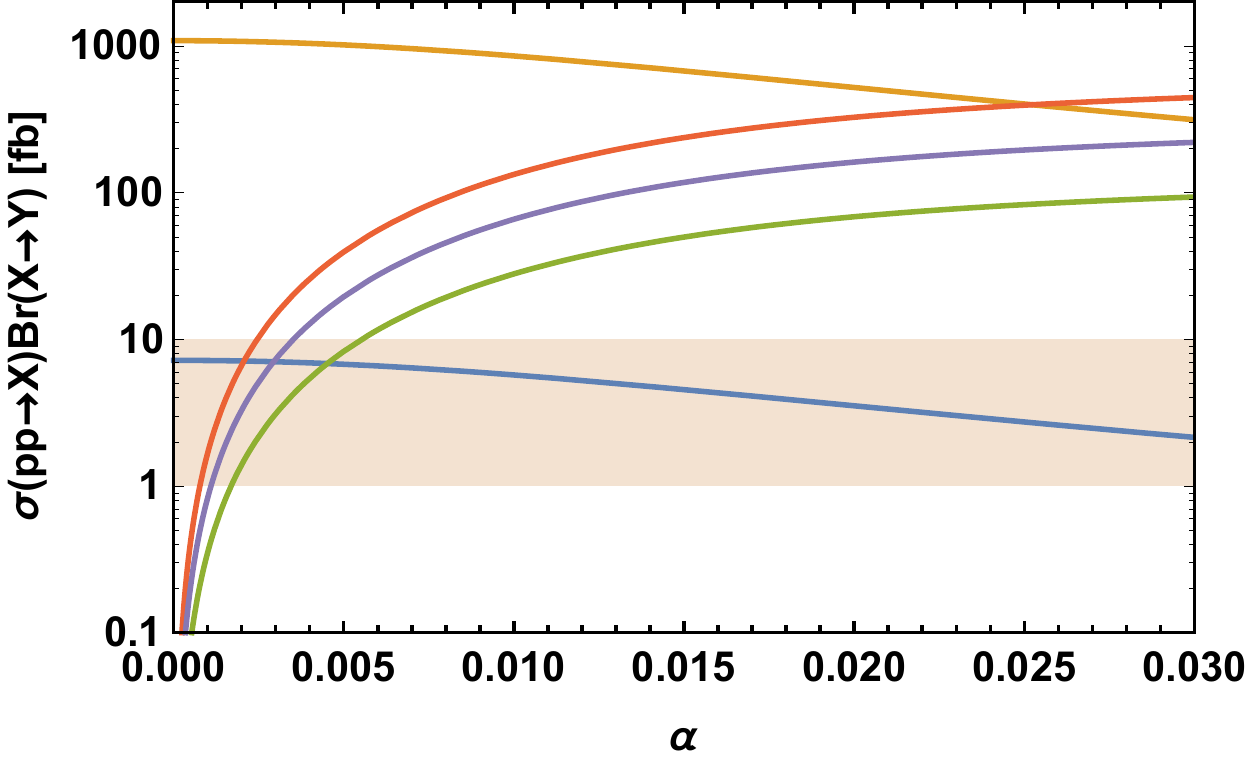}}
 \caption{The cross section times branching ratio for $p p \to X \to Y$ in fb for two leptoquarks models with $M_V = 1000$~GeV. The blue, orange, green, red, and purple curves correspond to $Y = \gamma\gamma,\, gg,\, t\bar{t},\, W^+W^-,\, \text{and } ZZ$ respectively. The shaded region corresponds to a cross section with the correct order of magnitude to explain the diphoton excess. For the model with quantum numbers $(\bar{3}, 1)_{-2/3}$, $\alpha > 0.012$ is ruled out by searches for resonances in the $W^+W^-$ and $ZZ$ channels.}
  \label{fig:vec2}
\end{figure}

Fig.~\ref{fig:vec} presents $\sigma(p p \to X)~\text{Br}( X \to \gamma \gamma)$ in fb as contours in the $M_V-\alpha$ plane. The left panel corresponds to a leptoquark with quantum numbers $(\bar{3}, 3)_{-2/3}$ and $\kappa = 0.10$ (and $g = 0$), whereas the leptoquark of the right panel has the quantum numbers $(\bar{3}, 1)_{-2/3}$ and $\kappa = 0.27$ (and again with $g = 0$). Of the five types of vector leptoquarks in Tab.~\ref{tab:vec}, these cases lead to the largest and smallest diphoton rate for a given value of $\kappa$ respectively. The orange and blue parameter space is ruled out by requiring the Higgs couplings to photons and gluons to be within 10\% of their SM values respectively. 
\begin{figure}
  \centering
  \subfloat[$(\bar{3}, 3)_{-2/3}$, $\kappa = 0.10$]{\includegraphics[width=0.48\textwidth]{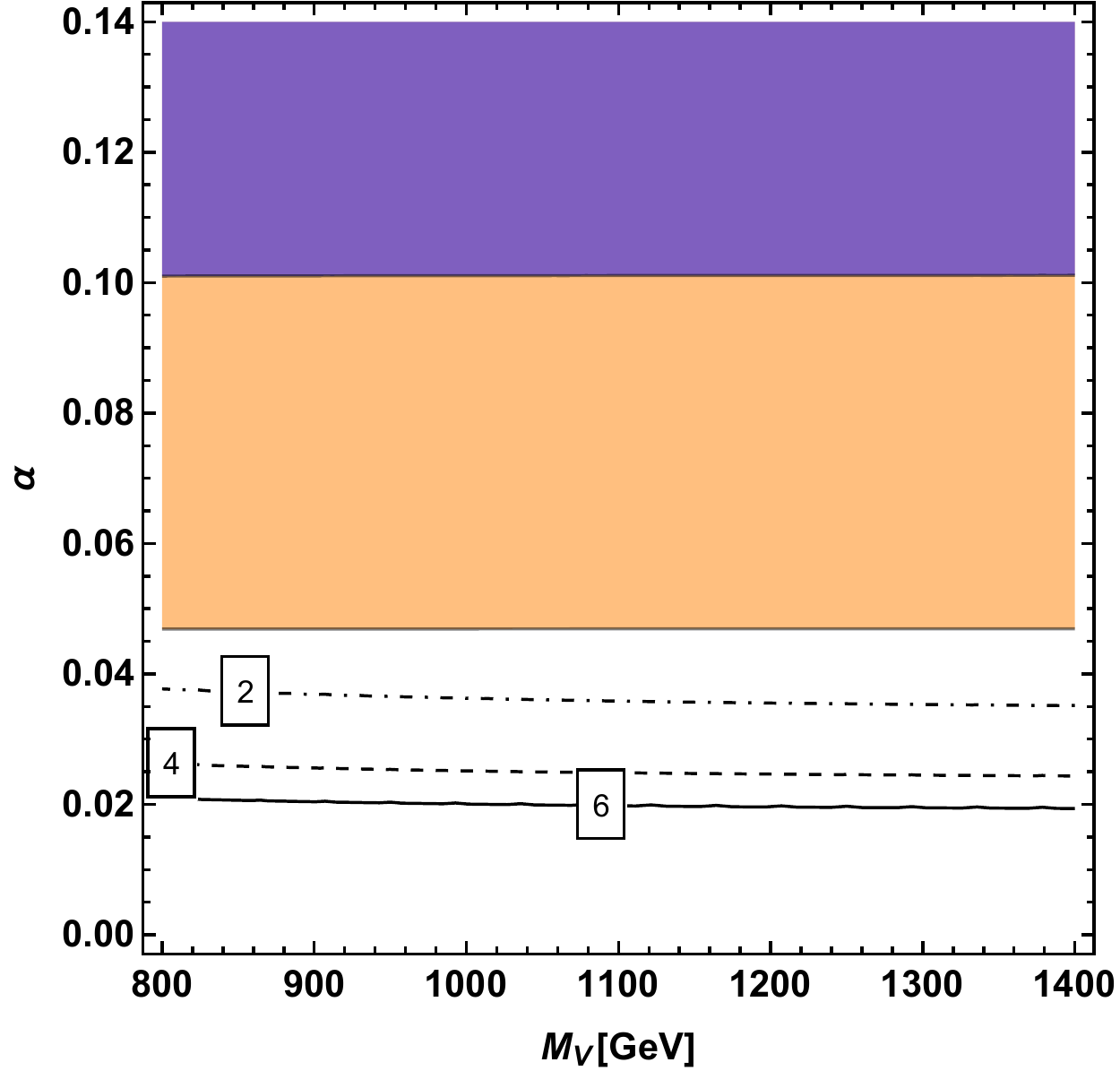}}\,
\subfloat[$(\bar{3}, 1)_{-2/3}$, $\kappa = 0.27$]{\includegraphics[width=0.48\textwidth]{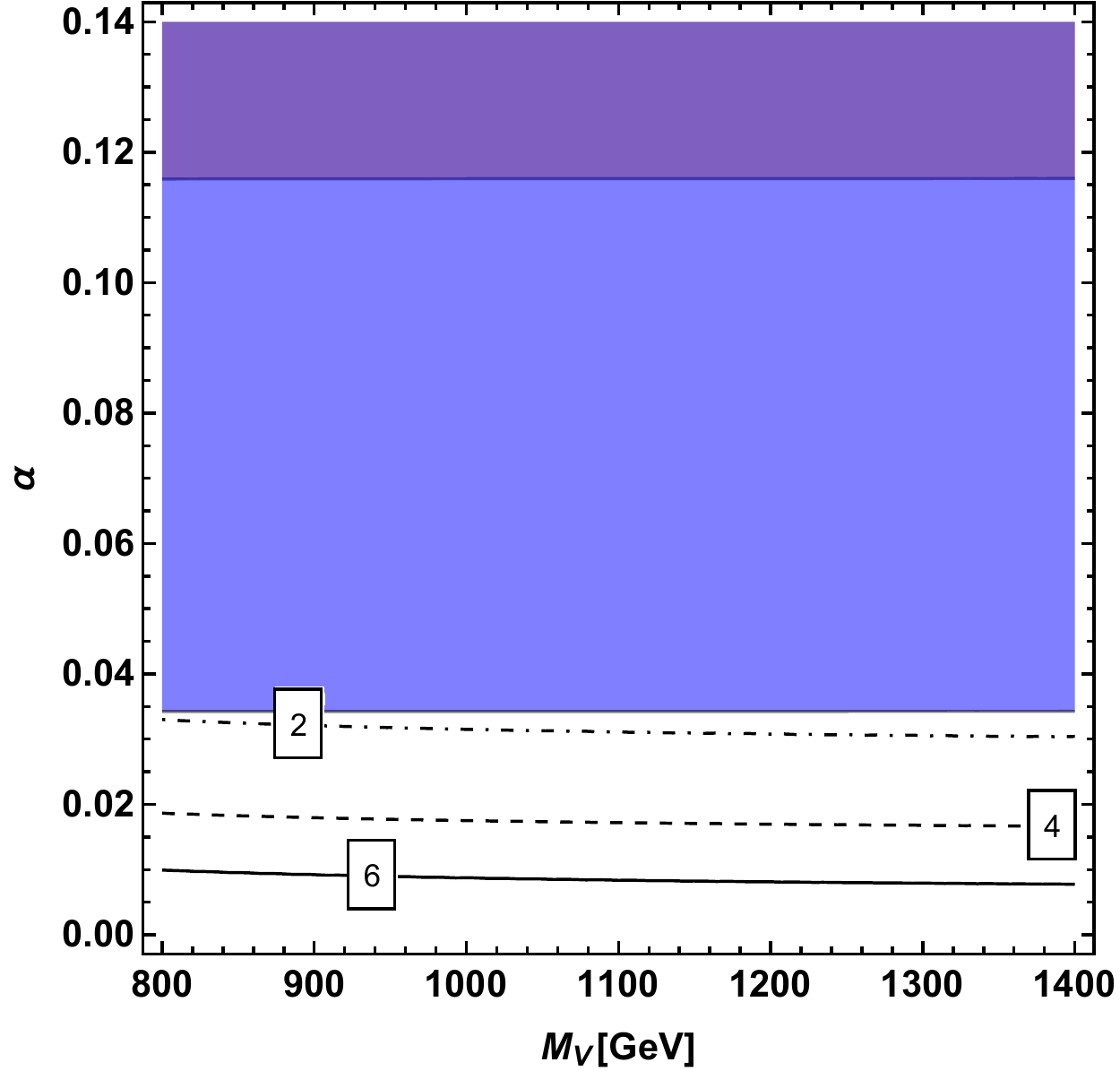}}
 \caption{The cross section times branching ratio for $p p \to X \to \gamma \gamma$ in fb for two leptoquarks models as contours in the $M_V - \alpha$ plane. The orange and blue parameter space is ruled out by requiring the Higgs couplings to photons and gluons to be within 10\% of their SM values respectively.}
  \label{fig:vec}
\end{figure}
Fig.~\ref{fig:vec2} shows cross section times branching ratio for $p p \to X \to Y$ in fb for two leptoquarks models with $M_V = 1000$~GeV. The blue, orange, green, red, and purple curves correspond to $Y = \gamma\gamma,\, gg,\, t\bar{t},\, W^+W^-,\, \text{and } ZZ$ respectively. The shaded region corresponds to a cross section with the correct order of magnitude to explain the diphoton excess. For the model with quantum numbers $(\bar{3}, 1)_{-2/3}$, $\alpha > 0.012$ is ruled out by searches for resonances in the $W^+W^-$ and $ZZ$ channels~\cite{CMS:2013ada, CMS:2015lda, Aad:2015kna,Aad:2015agg}. We find that these are the two most constraining direct searches for this model, rather than heavy quarks, taus, dijets, or $Z \gamma$. See e.g.~\cite{Knapen:2015dap, Franceschini:2015kwy} for more discussion of these bounds.

Figs.~\ref{fig:vec} and~\ref{fig:vec2} demonstrate that there is parameter space that gives the correct cross section for a putative 750 GeV resonance decaying to photons that is consistent with unitarity, measurements of the properties of the 125 GeV Higgs boson, and direct searches for resonances in other channels.

\section*{Acknowledgments}
We thank Riccardo Barbieri, Benjam\'{i}n Grinstein, Jorge Martin Camalich, Alfredo Urbano, and Patipan Uttayarat for helpful conversations. This work was supported in part by the MIUR-FIRB under grant no.~RBFR12H1MW.

\section*{References}
\bibliography{leptoquarks_v5}
\bibliographystyle{elsarticle-num}

\end{document}